\documentclass[fleqn,prd,onecolumn,10pt]{revtex4}

\usepackage{amsmath} \usepackage{graphicx} \usepackage{amsfonts}
\usepackage{array} \usepackage{amsthm}

\usepackage[dvips]{epsfig}
\usepackage{psfrag}

 \newcommand{\nn}{\nonumber}
\newcommand{\qd}{\qquad} 
\newcommand{\mn}{{\mu\nu}} 
 
\newcommand{\p}{\partial}

\newcommand{\0}{p_0}

\newcommand{\ra}{\rightarrow}
\newcommand{\wrho}{w_0\rho_{1}^2}
\newcommand{\rgb}{\mathcal{R}_{GB}^2}

\newcommand{\lge}{\mathcal{L}^{(E)}}
\newcommand{\lgs}{\mathcal{L}^{(str)}}
\newcommand{\ace}{\mathcal{S}^{(E)}}
\newcommand{\acs}{\mathcal{S}^{(str)}}
\begin{document}
\preprint{DTP-MSU/08-22}
\title{`Cusp' solutions in Gauss--Bonnet gravity}
\author{
    Evgeny A. Davydov}
\email{davydov@theor.jinr.ru} \affiliation{Joint Institute for
Nuclear  Research, Dubna, Moscow Region RU-141980}

\begin{abstract}
Einstein-dilaton-Gauss--Bonnet gravity is investigated on existence
of solutions with mild singularities, not shielded by the event
horizons. These still may have sense since presumably such
singularities will be smoothed by corrections to Einstein theory
from quantum gravity/string theory. We show that gravity with the
first-order correction, the Gauss--Bonnet term, gives rise to
special types of singularities, which we call `cusps', with $1/2$-th
and $1/3$-th powers in series expansion. The full space then can be
split onto several cusps with classically impenetrable borders,
and/or flat asymptotic.
\end{abstract}

\maketitle

\section{Gauss--Bonnet gravity}
Gauss--Bonnet term provides a non-trivial contribution to the
Einstein  equations only in  dimensions $D>4$, while in $D=4$
gravity one should include interaction with the dilaton field. The
solution without the GB correction is a standard Schwarzschild
metrics with constant dilaton (zero dilaton charge is ensured by the
`no-hair' theorem). Adding the GB term gives rise to a non-trivial
dilaton  configuration with two different types of asymptotically
flat solutions: the first  is the black hole, and the second
contains a naked singularity. Actually the BH solution  also
contains singularity, but this is shielded by the event horizon.

Solutions with naked singularities are usually not in a
favor,therefore subsequent investigations were produced only for the
black hole branch. But now let's turn back to our motivation of
adding a higher order curvature corrections to the action. It is
because of our belief in some fundamental quantum theory we add the
GB term. Therefore when we get a `singularity' this just means that
we moved outside the semi-classical limit with first-order
corrections, but the `real' quantum theory should deal with the
growing curvature and energy density in some way. This is why the
domain of initial conditions with high energy density should not be
forgotten in favor of the domain of initial conditions with the
event horizon. So in our work we will explore the family of
solutions with GB corrections in more details.

Again, one should mention that since the GB term is motivated by
string theory, the corresponding calculations should be produced
rather in the string frame than int the Einstein frame, and the
transformation of the Gauss--Bonnet term between these frames
generates a very complicated additional term. It was shown that not
all solutions known in  the Einstein frame are reproduced in the
string frame the string frame   \cite{Maeda:2009uy} in the context
of the dilaton black holes with Gauss--Bonnet corrections. For
example, the family of singular solutions, obtained by
\cite{Kanti:1995vq} as far as we know was not found in the string
frame.

Therefore in our investigation we will provide two parallel
calculations: for the usual Einstein-dilaton-Gauss--Bonnet action
(to extend known results)
\begin{equation}\label{eq:Sein1}
    \ace=\frac{1}{16\pi}\int \left( R-(\p_\mu \ln{S})^2/(2a^2)+\alpha
    S\rgb\right)\sqrt{-g}\,d^4x,
\end{equation}
and for the stringy version of the action
\begin{equation}\label{eq:Sstr1}
    \acs=\frac{1}{16\pi}\int \left(R+(\p_\mu \ln{S})^2/a^2+\alpha
    \rgb\right)S\sqrt{-g}\,d^4x.
\end{equation}
Note that in \cite{Maeda:2009uy} the first version was called a
`truncated' one. We shall use another notation: the first version
will be marked as ``(E)'' (EGBD), and the second version as
``(str)'' (SEGBD). Along with the Ricci scalar curvature $R$, it
contains the dilaton field $S=e^{2a\phi}$ and the Gauss--Bonnet term
$\rgb$. The Euler density of the GB term has the following form:
\begin{equation}\label{eq:GB1}
    \rgb=R^2-4R_{\mu\nu}R^{\mu\nu}+R_{\alpha\beta\mu\nu}R^{\alpha\beta\mu\nu}.
\end{equation}
The action is written in Plank units with $\hbar=c=G=1$, and it
contains two parameters: $a$ which is a dilaton coupling constant
and $\alpha$ which is a `correction' parameter in sense that GB
term is treated as a correction to the Einstein's action. Both
parameters are supposed to be positive.

The static spherically symmetric spacetime can be described by the
following metrics:
\begin{equation}\label{eq:g1}
    ds^2=-w(r)\sigma(r)^2dt^2+\frac{dr^2}{w(r)}+\rho(r)^2d\Omega_{2}^2,
\end{equation}
which contains three functions depending only on the radial
coordinate, but one of them can be gauged away by the rescaling of
$r$.

One can transform the action in the string frame (\ref{eq:Sstr1})
to the Einstein's frame by the conformal transformation
$g^{(str)}_{\mu\nu}= S^{-1}g^{(E)}_{\mu\nu}$:
\begin{eqnarray}
    \acs_{g}&=&\int R[g^{(E)}_\mn S^{-1}]S\sqrt{-g^{(E)}
    S^{-4}}\,d^4x =\nn\\
    &=&\ace_{g}-3/2\int (\p_\mu
    \ln{S})^2\sqrt{-g^{(E)}}\,d^4x.
\end{eqnarray}
In string theory one has $a=1$, therefore
$\mathcal{S}_{g}^{(str)}+\mathcal{S}_{d}^{(str)}=\mathcal{S}_{g}^{(E)}+\mathcal{S}_{d}^{(E)}$,
where $\mathcal{S}_d$ is a dilaton kinetic term. So in the case
$a=1$ the standard Einstein-dilaton action without GB corrections is
identical in both frames up to the conformal transformation. For
arbitrary $a$ the transformation from the string frame to the
Einstein frame produces the action with additional kinetic term.

The transformation of the GB term under the conformal
transformations is rather complicated. But we can just write the
GB-term in the non-transformed frame and substitute the metric
functions as $w^{(str)}=S w^{(E)},\:\sigma^{(str)}=S^{-1}
\sigma^{(E)},\:\rho^{(str)}=S^{-1/2} \rho^{(E)}$. This will allow us
to express both versions of the action in  one frame, where the
high-order curvature correction terms will be different.

Since the GB term is a total derivative:
\begin{equation}\label{GB2}
    \sqrt{-g}\rgb=\Lambda_{GB}',\quad \mbox{where}\quad
    \Lambda_{GB}=\frac{4}{\sigma}(w\sigma^2)'(w\rho'^2-1),
\end{equation}
it enters the lagrangian like $-\alpha S'\Lambda_{GB}$. The
expression above is correct for the $g^{(str)}_\mn$ metric functions
in the string frame and for the $g^{(E)}_\mn$ metric functions in
the Einstein frame. Therefore the GB part of the action
$\mathcal{S}^{(str)}$ in the Einstein frame will enter also like
$-\alpha S'\tilde{\Lambda}_{GB}$, but with the different expression
for the $\Lambda_{GB}$-term which now reads as
\begin{equation}\label{GB3}
    \tilde{\Lambda}_{GB}=\frac{4S}{\sigma}\left(\frac{w\sigma^2}{S}\right)'
    \left[Sw\left(\frac{\rho}{\sqrt{S}}\right)'^2-1\right],
\end{equation}
where the metric functions were substituted by the transformed ones.
This expression can be split onto the same $\Lambda_{GB}$-term when
the derivatives do not act on the dilatonic function and the terms
which are proportional to it's derivatives:
\begin{equation}
    \tilde{\Lambda}_{GB}=\Lambda_{GB}+\Delta\Lambda_{GB},\quad
    \mbox{where}\quad\Delta\Lambda_{GB}=\sum_{n=2}^4\Lambda_n(\ln{S})'^n.
\end{equation}
One can see that the difference between two forms of the GB term
is significant only when dilaton demonstrates the exponential
growth, and it vanishes if asymptotically $S\ra\mathrm{const}$.

Now we can obtain the one-dimensional effective EDGB lagrangian:
\begin{equation}\label{LE}
    \lge=\frac{\rho'(\sigma^2 w\rho)'}{2\sigma}-\frac{\sigma
    w\rho^2{S'}^2}{8a^2S ^2}-
    \frac{\alpha S'}{\sigma}(w\sigma^2)'(w\rho'^2-1).
\end{equation}
And the lagrangian of the stringy model reads as
\begin{equation}\label{LS}
    \lgs=S\left(\frac{\rho'(\sigma^2 w\rho)'}{2\sigma}+\frac{\sigma}{2}+\frac{\sigma
    w\rho^2S'^2}{4a^2S^2}\right)+\frac{S'(w\sigma^2\rho^4)'}{4\sigma\rho^2}-
    \frac{\alpha S'}{\sigma}(w\sigma^2)'(w\rho'^2-1).
\end{equation}

The first solution for the Einstein DGB system was obtained by Kanti
et. al. \cite{Kanti:1995vq} where they obtained the BH-branch of
solutions with horizon $r_h$ and another branch with an unshielded
singularity $r_x$ where square of the Riemann tensor diverges like
$(r-r_x)^{-1}$. Then, Alexeyev et. al. \cite{Alexeev:1996vs}
explored the behavior of the solutions under horizon $r_h$ and
revealed that then $r_s$ point becomes a ``turning point''
singularity from which two branches of solutions arise. The first
one goes outward from $r_s$ to $r_h$ and the second one moves
further into the black hole from $r_s$ to the singularity $r_x$. In
\cite{Melis:2005xt} the dynamical analysis was produced and it was
shown that the only regular solutions with flat asymptotic are the
black holes. In paper \cite{Maeda:2009uy} the authors reproduced the
BH solutions in the string frame and revealed the difference between
EDGB and SEDGB models, but they did not investigate non-BH solutions
in details, just mention that sometimes there is not a BH, but a
singularity.

\section{Cusp singularities}
We treat cusp as a mild singularity of the metrics, where the
dilaton  function and metric components are regular and
non-vanishing, but their second derivatives are singular.

In what follows we will use the shifted radial variable $x=r-r_s$
for convenience. In the vicinity of the singularity the metric and
dilaton functions can be written as expansions by powers of $x$:
\begin{equation}
  w = \sum\limits_{n=0}^\infty w_{n/z} x^{n/z}, \qd
  \rho = \sum\limits_{n=0}^\infty \rho_{n/z} x^{n/z}, \qd
  S = \sum\limits_{n=0}^\infty p_{n/z} x^{n/z}, \label{eq:series1}
\end{equation}
with $z=2,~3,\ldots$ parameterizing the order of singularity. First,
we will investigate the case $z=2$, which corresponds to the
turning point of the metrics. To provide the regularity of the
first derivatives we impose the condition
$w_{1/2}=\rho_{1/2}=S_{1/2}=0$. In this case the curvature has
$1/\sqrt{x}$ singularity at the cusp.

Substituting the expansions in to the equations of motion we find
that starting from the second order (i.e. $n/2\geq 2$) the
equations are linear by the $n/2$-th coefficients, expressing them
through the previous coefficients. Unfortunately, in the first
order (to be precious $3/2$ order) we obtained the system for the
$p_{3/2}$ and $q_{3/2}$ which contains polynomials of orders
higher than five. Therefore we are unable to provide explicit
expansions. Nevertheless, this system can be solved numerically
with any required precision when substituting quantities instead
of free parameters of the expansions.

The free parameters of the expansions are the parameters of the
configuration $(a,~\alpha)$, the dilaton value on the cusp surface,
$p_0$, the radius of the cusp, $\rho_0$, the angular deficit of the
cusp which can be expressed by $\rho_1$ and the time component of
the metrics, $w_0$. But the last two parameters can be combined into
one $u=\wrho$, while the remaining degree of freedom will be
responsible only for the scale transformations. In the case of  flat
asymptotical metrics  the GB term vanishes faster then the curvature
term, for which it is well known that the asymptotic should satisfy
the relation $w_\infty \rho_\infty'^2=1$, while separately
$w_\infty$ and $\rho_\infty'$ can be not equal to unity. Therefore
the parameter $u$ can be treated as a deviance  of the cusp metrics
from the flat asymptotic. Also the dilaton parameter $p_0$ can be
excluded by the appropriate rescaling of the dilaton function and
the GB parameter $\alpha\ra\alpha p_0$. In what follow we will set
$p_0=1$, so that the effective GB parameter $\alpha p_0$ will
coincide with the initial parameter $\alpha$ in the lagrangian.

\subsection{z=2 cusp}
\begin{figure}[p]
\hbox to\linewidth{\hss%
 \psfrag{1}{\huge{$x$}}
 \psfrag{2}{\huge{$w$}}
 \psfrag{3}{\huge{$\rho'$}}
 \psfrag{4}{\huge{$S$}}
    \resizebox{9.5cm}{6.5cm}{\includegraphics{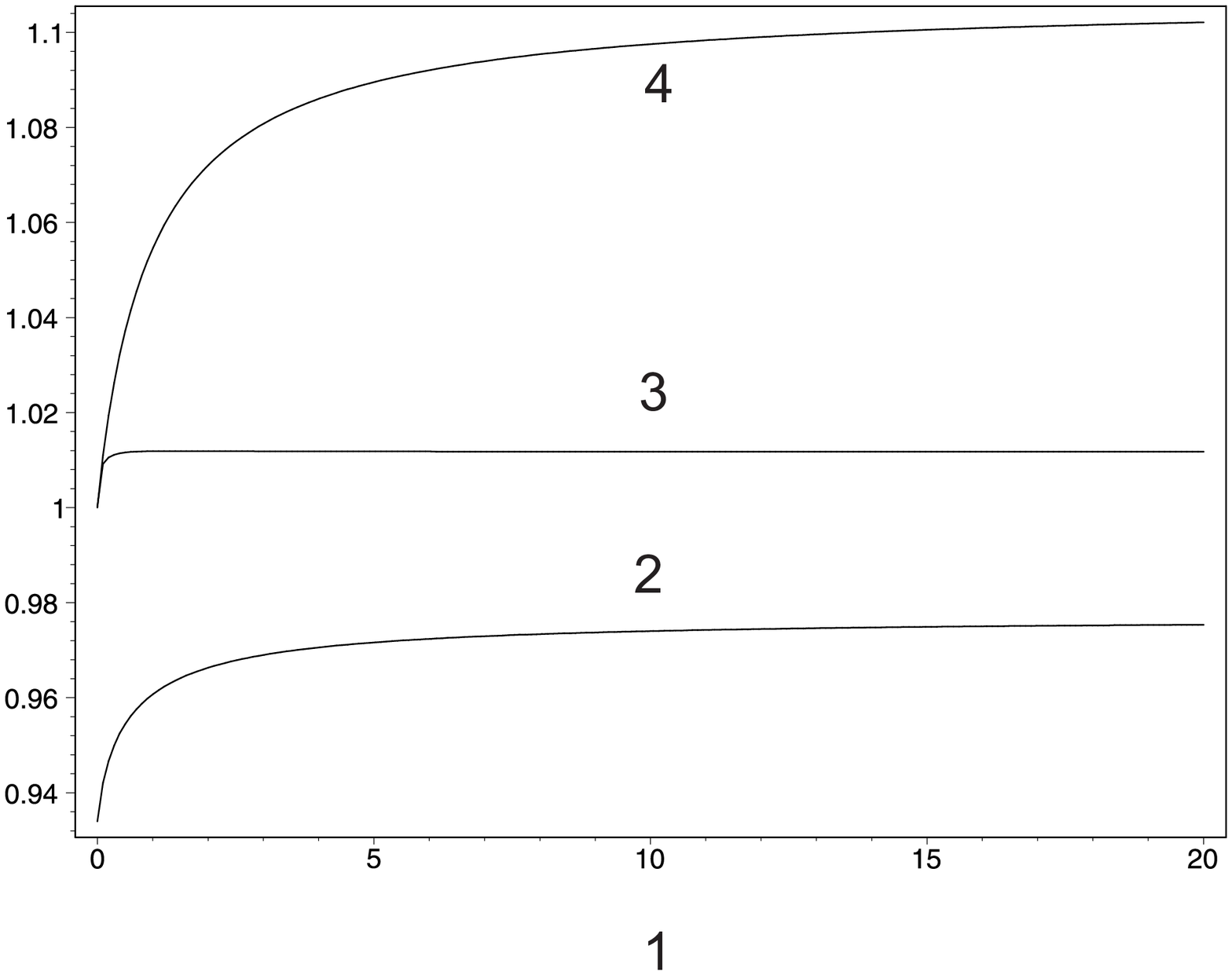}}
\hss}\caption{\small $z=2$: from cusp to Minkowski asymptotic.}\label{fig:z2flat}
\end{figure}

\begin{figure}[p]
\hbox to\linewidth{\hss%
 \psfrag{1}{\huge{$x$}}
 \psfrag{2}{\huge{$w/x^2$}}
 \psfrag{3}{\huge{$S/\sqrt{x}$}}
 \psfrag{4}{\LARGE
 {$\rho/\ln{x}$}}
    \resizebox{9.5cm}{6.5cm}{\includegraphics{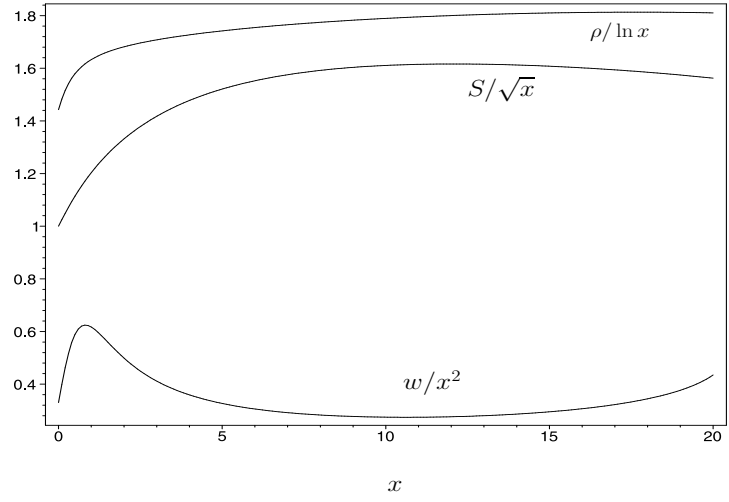}}
\hss}\caption{\small $z=2$: from cusp to cusp.}\label{fig:z2z2}
\end{figure}

The most interesting solutions are those  which interpolate between
the cusp and the flat asymptotic (Fig.~\ref{fig:z2flat}), and
numerical computations confirmed their existence. Other solutions
interpolate between two cusps of different radii
(Fig.~\ref{fig:z2z2}). The actual parameter space is
four-dimensional $(a,\,\alpha,\,\rho_0,\,u)$ but it can be
schematically described in the following way. The key role in
producing the asymptotically flat solutions belongs to the parameter
$u$. It appeared that the deviance from the asymptotic should be
small enough, but non-vanishing. There is a forbidden gap around
$u=1$, and two permitted bands on each side of it. The decrease of
$a,\,\alpha$ and the increase of $\rho_0$ causes the stretching of
the allowed bands, while the forbidden gap also enlarges above $u=1$
and slightly modifies below $u=1$. Conversely, the increase of
$a,\,\alpha$ and the decrease of $\rho_0$ causes the contraction of
the both permitted bands and the forbidden gap above $u=1$. To be
precise, these bands are not solid, they possess stripy structure
with narrow forbidden gaps. But the detailing seems to be abundant
for practical purposes.

One can conclude that all conditions which stimulate existence of
asymptotically flat solutions can be understood as relative decrease
of the contribution of Gauss--Bonnet corrections in comparison with
the classical Einstein-dilaton gravity terms. This can  be seen from
the table which follow.  Data in each table starts from the point
$a=\alpha=\rho_0=1$ and then one of the parameters varies in the
direction which enlarges the allowed bands for $u$.

\begin{center}
\begin{tabular}{||l|cc|cc||}
  \hline
  $(a,~\alpha,~\rho_0)$ & allowed & band & allowed & band \\
  \hline
  (1,~1,~1) & 0.93 & 0.99 & 1.02 & 1.11\\
  (0.5,~1,~1) & 0.77 & 0.99 & 1.09 & 1.22\\
  (0.33,~1,~1) & 0.65 & 0.98 & 1.1 & 1.34\\
  (1,~0.5,~1) & 0.89 & 0.98 & 1.06 & 1.23 \\
  (1,~0.33,~1) & 0.75 & 0.94 & 1.11 & 1.34 \\
  (1,~1,~2) & 0.46 & 0.92 & 1.16 & 1.55 \\
  (1,~1,~3) & 0.6 & 0.86 & 1.54 & 2.5 \\
  \hline
\end{tabular}
\end{center}

One can compute the ADM mass and the dilaton charge of the
configuration using the asymptotic obtained numerically. It appears
that the dilaton charge decreases and the ADM mass increases when
the parameter $u$ deviates from unity. The dilaton charge is
negative, while the mass is positive on the lower band and negative
for on upper band.

There is another interesting branch of solutions for the
considerable deviance from the asymptotic with $u\sim 0.6$. With
growing $x$ one obtains solutions with the metrics reaching the flat
asymptotic, while the dilaton function grows exponentially. Still
these solutions are singular, but the point of singularity moves to
infinity with the decrease of $\alpha$. Actually the plot of the
dilaton function looks like that of the scale factor in the
inflationary scenario with fast acceleration period turning then
into slow deceleration. With the significant decrease of $\alpha$
this branch disappears since for sufficiently small  $\alpha$ the
lower allowed band of regular solutions reaches the point $u\sim
0.6$.

\subsection{z=3 cusp}

One can choose the value $z=3$, i.e. the expansions near cusp will
contain powers of $n/3$. In this case the condition of regularity
of the first derivatives does not hold, but there is a
$w_{2/3}x^{1/3}$ term. The scalar curvature possess stronger
singularity $x^{-4/3}$ instead of the square root singularity as
it was for $z=2$.

Unlike the previous case, the coefficients of the expansions can be
found explicitly, but they are rather complicated. It turned out
that there are two branches of the expansions depending on whether
$\rho_{4/3}$ vanishes or not. Moreover, each branch splits into two
other branches which are identical up to the transformation
$x^{n/z}\ra (-1)^n x^{n/z}$, which defines the `direction' where the
solution spreads. The parameter space contains five parameters, but
again $p_0$ and $w_0$ can be scaled away. So the effective parameter
space $(a,~\alpha,~\rho_0)$ is three-dimensional, and there is no
independent parameter like $u$ in previous case.

\begin{figure}[p]
\hbox to\linewidth{\hss%
 \psfrag{1}{\huge{$x$}}
 \psfrag{2}{\huge{$\rho'$}}
 \psfrag{3}{\huge{$(\ln{S})'$}}
 \psfrag{4}{\huge{$w$}}
    \resizebox{9.5cm}{6.5cm}{\includegraphics{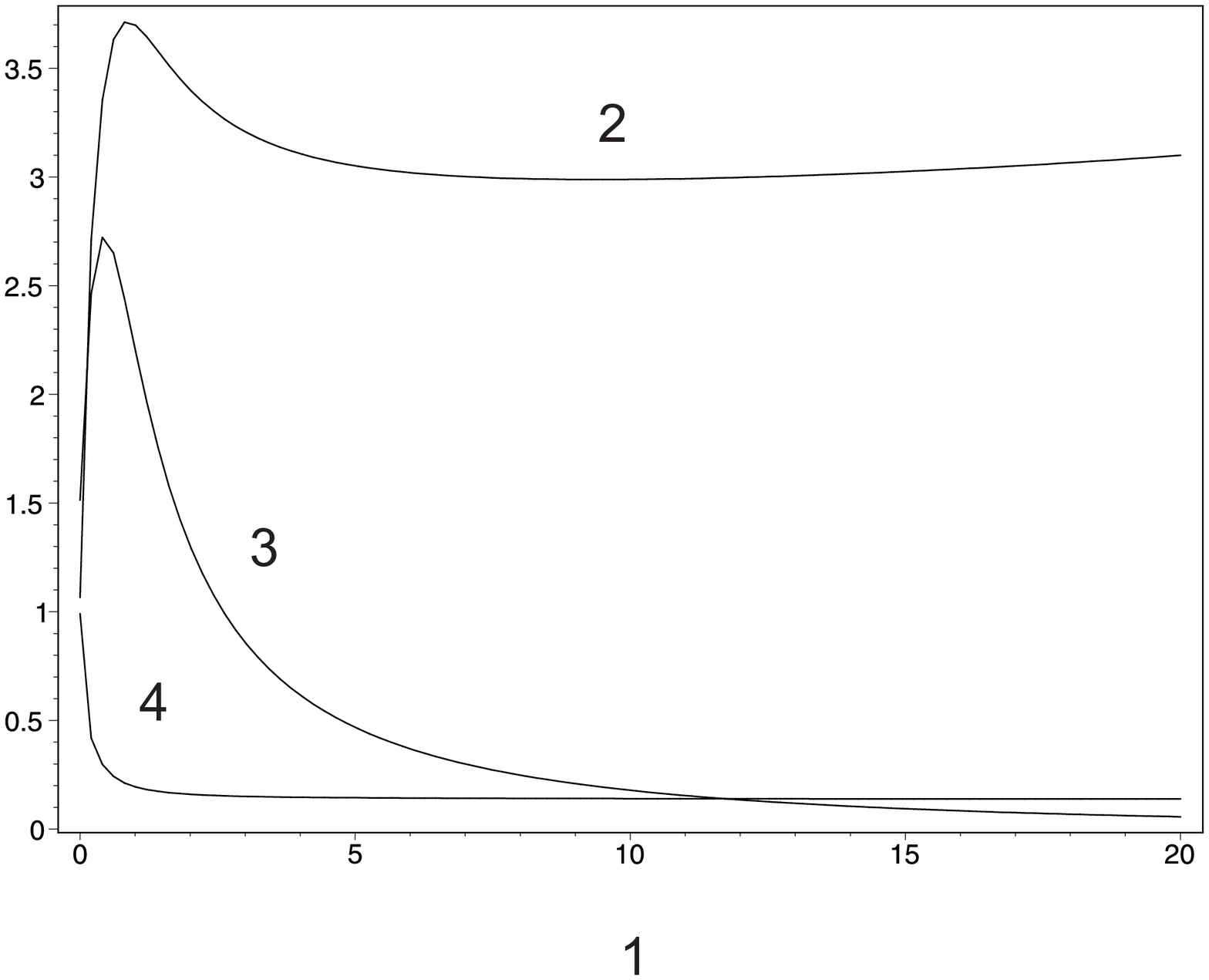}}
\hss}\caption{\small $z=3$: from cusp to cusp $z=2$ through the flat middle area.}\label{fig:z3}
\end{figure}

Surprisingly, we did not found any regular solution with flat
asymptotic numerically. Each solution ends by the square-root cusp
considered in previous section as shown on (Fig.~\ref{fig:z3}).
Nevertheless, there exist the same branch of solutions with the
exponentially growing dilaton function when $\alpha$ decreases. But
now the absence of regular solutions does not bound this branch from
below. So with $\alpha\ra\0$ on has $S\ra\infty$ and $w,~\rho'\ra
\mathrm{const}$ what gives us the flat metrics. The point of
singularity moves to infinity. Mention that there is no cusp for
$\alpha=0$, since the terms with non-integer powers of $x$ vanish in
the expansions both for $z=2$ and $z=3$.

\section{Conclusion}

We investigated the class of singular solutions to Gauss--Bonnet
dilaton  gravity system. First, we mentioned already known singular
solution with $1/2$-th powers expansion near singularity. Such
solutions can provide flat asymptotic, but also they can demonstrate
from-cusp-to-cusp behavior, where the entire space can be split onto
several regions by cusps which are classically impenetrable. Then we
showed the existence of a new singular solution to Gauss--Bonnet
dilatonic gravity, containing $1/3$-th powers near the singular
point. It can not produce flat asymptotic and always spreads from
one singularity to another. But it improves the inner region in the
sense that one has flat space between cusps.

Still there is an open question if there are other cusps, probably
even with not a power-series expansions. Such nonlinear differential
equations can provide a very diverse picture. We investigated
$1/4$- and $1/5$-th powers and found no cusp solutions. Another
question which we explored was the difference between classical and
string versions of dilatonic gravity with curvature corrections. It
appeared that qualitatively the picture is almost the same.

One can ask about applications of cusp solutions. Some of them can
be interesting in view of the brane-world scenarios, where the
singularity is moved to the bulk
\cite{Feinstein:2001xs,Kunze:2001ji,BouhmadiLopez:2008hk}. Finally,
if the black holes were treated as some unnatural solutions for many
years, and now we have several dozens of candidates to them,
probably another even more strange configurations will arise in
future astrophysics.

This work was supported by Dynasty foundation and the RFBR grant
08-02-01398-a.

\begin{thebibliography}{9}


\bibitem{Maeda:2009uy}
  K.~i.~Maeda, N.~Ohta and Y.~Sasagawa,
  ``Black Hole Solutions in String Theory with Gauss-Bonnet Curvature
  Correction,''
  Phys.\ Rev.\  D {\bf 80} (2009) 104032
  [arXiv:hep-th/0908.4151].



\bibitem{Kanti:1995vq}
  P.~Kanti, N.~E.~Mavromatos, J.~Rizos, K.~Tamvakis and E.~Winstanley,
  ``Dilatonic Black Holes in Higher Curvature String Gravity,''
  Phys.\ Rev.\  D {\bf 54}, 5049 (1996)
  [arXiv:hep-th/9511071].

\bibitem{Alexeev:1996vs}
  S.~O.~Alexeev and M.~V.~Pomazanov,
  ``Black hole solutions with dilatonic hair in higher curvature gravity,''
  Phys.\ Rev.\  D {\bf 55}, 2110 (1997)
  [arXiv:hep-th/9605106].


\bibitem{Melis:2005xt}
  M.~Melis and S.~Mignemi,
  ``Global properties of dilatonic Gauss-Bonnet black holes,''
  Class.\ Quant.\ Grav.\  {\bf 22} (2005) 3169
  [arXiv:gr-qc/0501087].



\bibitem{Feinstein:2001xs}
  A.~Feinstein, K.~E.~Kunze and M.~A.~Vazquez-Mozo,
  ``Curved dilatonic brane worlds,''
  Phys.\ Rev.\  D {\bf 64}, 084015 (2001)
  [arXiv:hep-th/0105182].


\bibitem{Kunze:2001ji}
  K.~E.~Kunze and M.~A.~Vazquez-Mozo,
  ``Quintessential brane cosmology,''
  Phys.\ Rev.\  D {\bf 65} (2002) 044002
  [arXiv:hep-th/0109038].

\bibitem{BouhmadiLopez:2008hk}
  M.~Bouhmadi-Lopez,
  ``Exploring the dark side of the Universe in a dilatonic brane-world
  scenario,''
  arXiv:0811.4069 [hep-th].





\end{thebibliography}
\end{document}